\documentclass{article}
\usepackage[utf8]{inputenc}
\usepackage{mathtools,amssymb}
\usepackage[usenames,dvipsnames,table]{xcolor}
\usepackage[titletoc]{appendix}
\usepackage{mathtools,amssymb}
\usepackage[makeroom]{cancel}
\usepackage{graphicx}
\usepackage{epstopdf}
\usepackage{mathrsfs}
\usepackage{enumitem}
\usepackage{stackrel}
\usepackage{float}
\usepackage{booktabs}
\usepackage[normalem]{ulem}  
\usepackage{cancel} 
\usepackage{hyperref}
\usepackage[draft,color,final]{showkeys} 
\usepackage{longtable}
\usepackage{physics}
\usepackage{empheq}
\usepackage{cite} 
\usepackage{comment}
\usepackage{tikz}
\usepackage{braket}
\usepackage{fullpage}
\usetikzlibrary{decorations.pathreplacing}

\DeclareRobustCommand{\rchi}{{\mathpalette\irchi\relax}}
\newcommand{\irchi}[2]{\raisebox{\depth}{$#1\chi$}}

\DeclareMathAlphabet{\mathpzc}{OT1}{pzc}{m}{it}

\allowdisplaybreaks
\allowbreak

\definecolor{refkey}{rgb}{0.9451,0.2706,0.4941}
\definecolor{labelkey}{rgb}{0.9451,0.2706,0.4941}
\renewcommand*\showkeyslabelformat[1]{%
  \fbox{\parbox[t]{0.8\marginparwidth}{\raggedright\normalfont\scriptsize\url{#1}}}}

\usepackage{etoolbox}
\makeatletter
\patchcmd{\hyper@makecurrent}{%
    \ifx\Hy@param\Hy@chapterstring
    \let\Hy@param\Hy@chapapp
    \fi
}{%
    \iftoggle{inappendix}{
	\@checkappendixparam{chapter}%
	\@checkappendixparam{section}%
	\@checkappendixparam{subsection}%
	\@checkappendixparam{subsubsection}%
	\@checkappendixparam{paragraph}%
	\@checkappendixparam{subparagraph}%
    }{}%
}{}{ \errmessage{failed to patch}}

\newcommand*{\@checkappendixparam}[1]{%
	\def\@checkappendixparamtmp{#1}%
	\ifx\Hy@param\@checkappendixparamtmp
	\let\Hy@param\Hy@appendixstring
	\fi
}
\makeatletter

\newtoggle{inappendix}
\togglefalse{inappendix}

\apptocmd{\appendix}{\toggletrue{inappendix}}{}{\errmessage{failed to patch}}
\apptocmd{\subappendices}{\toggletrue{inappendix}}{}{\errmessage{failed to patch}}

\begin{document}

\newcommand{\f}{\frac}
\newcommand{\z}{\zeta}
\newcommand{\del}{\partial}
\def\eq#1{(\ref{#1})}
\newcommand{\R}{\textcolor{red}}
\newcommand{\B}{\textcolor{blue}}
\newcommand{\tx}{\text{x}}

\DeclareRobustCommand{\rchi}{{\mathpalette\irchi\relax}}

\hypersetup{pageanchor=false}

\begin{titlepage}
			
\vspace{.4cm}
\begin{center}
\noindent{\Large \bf{Boundary Liouville Conformal Field Theory in Four Dimensions}}\\
\vspace{1cm} 
Adwait Gaikwad
\footnote{ adwait@tauex.tau.ac.il},
Amitay C. Kislev
\footnote{ amitaykislevtau@gmail.com},
 Tom Levy
\footnote{toml@mail.tau.ac.il}
 and Yaron Oz
\footnote{yaronoz@tauex.tau.ac.il}
\vspace{.5cm}
\begin{center}
{\it  School of Physics and Astronomy,}\\
{\it Tel-Aviv University, Tel-Aviv 69978, Israel}
\end{center}

				
\end{center}
\vspace{1cm}			
\begin{abstract}

Higher dimensional Euclidean Liouville conformal field theories (LCFTs)
consist of a log-correlated real scalar field with a background charge and an exponential potential. 
We analyse the LCFT on a four-dimensional manifold with a boundary. We extend  to the boundary, the conformally covariant GJMS operator and 
the ${\cal Q}$-curvature term in the LCFT action and classify the conformal boundary conditions.
Working on a flat space with plate boundary, we 
calculate the dimensions of the boundary conformal primary operators, the two- and three-point functions of the displacement
operator and the boundary conformal anomaly coefficients.

\end{abstract}

\end{titlepage}
		
\pagenumbering{roman}
\tableofcontents
\pagenumbering{arabic}
\setcounter{page}{1}

\newpage 
\section{Introduction}

Two-dimensional Liouville conformal field theory (LCFT) has been proposed as
a description of two-dimensional quantum gravity \cite{Polyakov:1981rd}.
A higher dimensional Euclidean generalization of LCFT
\cite{Levy:2018bdc} has been introduced in 
\cite{Oz:2017ihc,oz2020turbulence} as part of a Euclidean field theory description of steady state incompressible fluid turbulence.
In this framework, the Liouville scalar field is the Nambu-Goldstone boson associated with a spontaneous symmetry breaking (SSB)
due to the energy flux of the separate space and time scale symmetries of turbulence in the inertial range of scales.
This SSB to Kolmogorov linear scaling \cite{Oz:2017ihc,oz2020turbulence} provides an analytical 
formula for the universal anomalous scaling exponents of turbulence \cite{Eling:2015mxa,Oz:2017ihc,oz2020turbulence}
by a generalization of KPZ scaling in two-dimensional quantum gravity \cite{Knizhnik:1988ak,DISTLER1989509}.
Mathematical aspects of higher dimensional LCFT have been addressed in
\cite{furlan2018some,schiavo2021conformally,cercle2022liouville}.
Higher dimensional LCFT appears
in the study of anomalies \cite{nakayama2018realization,nakagawa2020c}, in CFT complexity 
\cite{Boruch_2021,camargo2022q,chagnet2022complexity}, in Anderson transistion \cite{padayasi2023conformal}
and potentially in RG flows in tensor models \cite{bosschaert2022chaotic,jepsen2023rg}.
LCFT in dimensions higher than two is defined by a higher derivative action (see \cite{deUrries:1998obu,Osborn:2016bev,Stergiou:2022qqj,Chalabi:2022qit} for work on higher derivative field theories).
In odd number of dimensions the LCFT is non-local \cite{kislev2022odd} (see work on non-local field theories
\cite{heydeman2020renormalization,heydeman2023polyakov}).
The four-dimensional LCFT has $N=1$ and $N=2$ superconformal generalizations
studied in  \cite{Levy_2018,levy2019n} (see \cite{kuzenko2020non} for higher derivative supersymmetry).

The action of LCFT on a  $d$-dimensional manifold $\mathcal{M}$ reads:
\begin{equation}\label{eq:bulkac}
S_L [\phi,g] = \frac{d}{2 \Omega_d (d-1)!} \int_\mathcal{M} d^dx \sqrt{g} \left(\phi \mathcal{P}_g \phi + 2 Q \mathcal{Q}_g \phi + \f{2}{d}\Omega_d (d-1)!\ \mu e^{d b\phi}\right) \ .
\end{equation}
$\phi$ is the Liouville log-correlated  scalar field, $\Omega_d = 2\pi^{(d+1)/2}/\Gamma[(d+1)/2]$ is the surface volume of the d-dimensional sphere $S^d$, and the dimensionless parameters are $Q$, $\mu$ and $b$.  $\mathcal{P}_g$ is the conformally covariant GJMS operator \cite{10.1112/jlms/s2-46.3.557}:
\begin{equation}
    \mathcal{P}_g = (-\Delta)^{d/2} + \cdots \ ,
\end{equation}
where $x^\mu$ are local coordinates on  $\mathcal{M}$, $\Delta = g^{\mu\nu}\nabla_\mu \nabla_\nu$ with $\nabla_\mu$ being the covariant derivative and $\cdots$ are subleading derivative terms.  The $\mathcal{Q}$-curvature scalar \cite{0e8e2d4b-05f6-3a9e-92d4-e3c346d2b40e}, $\mathcal{Q}_g$, is:
 \begin{equation}
     \mathcal{Q}_g = \f{1}{(d-2)}\ (-\Delta)^{(d/2)-1} R + \cdots \ ,
 \end{equation}
where $R$ is the Ricci scalar of $\mathcal{M}$. 
The integral of $\mathcal{Q}_g$ is a conformal invariant of the manifold $\mathcal{M}$.

Under a $d$-dimensional Weyl transformation:
\begin{subequations}\label{eq:phiweyl}
\begin{align}
g_{\mu \nu} & \rightarrow e^{2\sigma} g_{\mu \nu} \, , \\
\phi & \rightarrow \phi -Q \sigma \, , \\
\mathcal{P}_g & \rightarrow e^{-d\sigma} \mathcal{P}_g  \, , \\
\mathcal{Q}_g & \rightarrow e^{-d\sigma} (\mathcal{Q}_g + \mathcal{P}_g\sigma)  \ , \end{align}
\end{subequations}
the action \eq{eq:bulkac} transforms as 
\begin{equation}\label{eq:actran}
    S_{CG}[\phi - Q\sigma, e^{2\sigma} g] = S_{CG}[\phi,g] - S_{CG}[Q\sigma,g] \ ,
\end{equation}
where $S_{CG} \equiv S_{L}(\mu=0)$  is the higher dimensional Coulomb gas theory. The complete Liouville action \eq{eq:bulkac} transforms as \eq{eq:actran} when we fix $Q = \f{1}{b}$. Thus, the action is Weyl invariant up to a term that does not affect the classical field equations. Semi-classically, the
critical points of the even-dimensional LCFT functional answer the problem of Uniformization,
i.e. does every compact even-dimensional manifold carry a conformal metric with constant negative ${\cal Q}$-curvature.

 We will work in four dimensions, where the GJMS (Paneitz) operator and the ${\cal Q}$-curvature read:
\begin{equation}
\mathcal{P}_g = \Delta^2 + 2 \nabla_\mu \left(R^{\mu\nu} - \frac{1}{3} g^{\mu\nu} R\right) \nabla_\nu \, , \quad \quad \mathcal{Q}_g = -\frac{1}{6} \left(\Delta R + 3 R_{\mu\nu} R^{\mu\nu} - R^2\right) \ , 
\label{4dpq}
\end{equation}
with $R_{\mu\nu}$ being the Ricci tensor of $\mathcal{M}$.

The aim of this work is to study the four-dimensional LCFT on a manifold with a boundary.
We will extend the LCFT action  \eq{eq:bulkac} to the boundary by extending the GJMS operator and the ${\cal Q}$-curvature
\eq{4dpq}, and we will classify  
 the consistent conformal boundary conditions. 
 Working on the ball and on the conformally equivalent flat space with a plate boundary, we will
calculate the dimensions of the boundary conformal primary operators, the two- and three-point functions of the displacement
operator and the boundary conformal anomaly coefficients. 
The paper is organized as follows. In section \ref{sec:2}, we introduce conformal geometry objects that  are needed 
in order to extend to the boundary the GJMS operator and the ${\cal Q}$-curvature, as well as to specify the conformal boundary conditions.
We construct the complete bulk plus boundary action and its Weyl transformation analogous to \eq{eq:actran}. In subsection \ref{subs:bdyspec} we work on the ball, or equivalently on a flat space with a plate boundary. We define the boundary vertex operators, derive their conformal dimensions
and analyse the restrictions on the boundary Liouville interaction. In subsection \ref{sec:BC}, we classify the set of boundary conditions required for a consistent variation of the action. In section \ref{sec:3}, we construct the displacement operator for the different boundary conditions  and compute the B-type boundary anomaly coefficients. In section \ref{sec:5} we conclude with an outlook.
Details of the displacement operator and the conformal mapping between the ball and a flat space with a plate boundary
are given in the appendices.

\section{Boundary Liouville CFT in Four Dimensions}\label{sec:2}
Let $\mathcal{M}$ be a four-dimensional manifold with a boundary $\partial \mathcal{M}$. We will use Greek alphabet to index the bulk coordinates $x^\mu$, and Roman alphabet to index the boundary coordinates $x^i$. We will also denote the boundary position vector as $\mathbf{x}$. The boundary is defined by the embedding map $X^{\mu}(\mathbf{x})$. The coordinates are dimensionful and the metric is dimensionless. We denote the normal vector to the boundary by $n^{\mu}$, and
the normal derivative as $\partial_{n}\equiv n^{\mu}\partial_{\mu}$.
The induced metric on the boundary (first fundamental form) is $\hat{g}_{ij} = \frac{\partial X^{\mu}}{\partial \mathrm{x}^{i}} \frac{\partial X^{\nu}}{\partial \mathrm{x}^{j}}\ g_{\mu\nu}$, where $g_{\mu\nu}$ is the bulk metric. The extrinsic curvature (second fundamental form) is $K_{ij} = \frac{\partial X^{\mu}}{\partial \mathrm{x}^{i}} \frac{\partial X^{\nu}}{\partial \mathrm{x}^{j}} \nabla_{\mu}n_{\nu}$. We define the traceless part of the extrinsic curvature 
as $\mathring{K}_{ij} = K_{ij}-\frac{K}{3}g_{ij}$, where $K$ is its trace. For correlation function computations, we will consider the upper half space $\mathbb{H}\mathbb{R}^4$, which has a flat metric in the bulk and a plate boundary,  i.e. $y=0$, $x^\mu = (y,\mathbf{x})$.

\subsection{Boundary Conformal Primary Operators}
A set of conformally covariant boundary operators associated to the Paneitz operator (GJMS operator in four dimensions) $\mathcal{P}_g$ with a general background and boundary was presented in \cite{case2015boundary} (also see \cite{case2018boundary}).  These  boundary operators read:
\begin{subequations}\label{eq:B}
\begin{align}
\mathcal{B}_{g,0}\phi = \Phi^{(0)} &= \phi|_{\del \mathcal{M}},  \\
\mathcal{B}_{g,1}\phi = \Phi^{(1)} &= \del_{n}\phi|_{\del \mathcal{M}},  \\
\mathcal{B}_{g,2}\phi = \Phi^{(2)}  &= \left(-\hat{\Delta}+n^{\mu}n^{\nu}\nabla_{\mu}\del_{\nu}+\frac{1}{3}K\partial_{n}\right)\phi\bigg|_{\del \mathcal{M}}, \\
\mathcal{B}_{g,3}\phi = \Phi^{(3)}  &= \left(-\del_{n}\Delta -2\hat{\Delta}\del_{n}+2\mathring{K}^{ij}\hat{\nabla}_{i}\del_{j}-\frac{2}{3}K\hat{\Delta}+\frac{2}{3}\del^{i}K\del_{i}+\mathcal{S}_{g,2}\del_{n} \right) \phi\bigg|_{\del \mathcal{M}} \ ,
\end{align}
\end{subequations}
where
\begin{equation}
\mathcal{S}_{g,2} = \frac{1}{2}\hat{R}+\frac{1}{6}R-R_{\mu\nu}n^{\mu}n^{\nu}-\frac{1}{3}K^2+\frac{1}{2}\mathring{K}^{ij}\mathring{K}_{ij} \ ,
\end{equation}
and the quantities with hat belong purely to the boundary. In the case of a plate boundary, it is easy to see the construction of these boundary conformal primaries. The details are given in \cite{Chalabi:2022qit}, here we describe the logic of the derivation. Out of all the generators of the conformal group $SO(d,2)$, $P_n$ (translations normal to the boundary), $K_n$ (boosts normal the boundary), and $M_{\mu n}$ (rotations in a plane with one axis along and the other transverse to the boundary), where $n$ denotes the direction normal to the boundary, are broken. New boundary primary operators can be build by acting on a bulk primary operator by powers of $P_n$ and ensuring that the state is annihilated by all $K_i$ and not necessarily by $K_n$.\\
We are interested in quantities which have nice Weyl transformations for the construction of the boundary action. The quantities which are useful to us along with \eq{eq:B} are scalar invariants called T-curvatures: 
\begin{subequations}\label{eq:T}
 \begin{align}
\mathcal{T}_{g,1} &= \frac{1}{3}K \, , \\
\mathcal{T}_{g,2} &=  \frac{1}{4}\hat{R}+\frac{1}{12}R-\frac{1}{2}R_{\mu\nu}n^{\mu}n^{\nu}+\frac{1}{18}K^2 \, , \\
\mathcal{T}_{g,3} &= \frac{1}{6} \partial_n R - \frac{2}{3} \hat{\Delta} K - 2 \mathring{K}^{ij}  \hat{R}_{ij} + \frac{1}{3} K \hat{R} + \frac{1}{3} K\mathring{K}^{ij} \mathring{K}_{ij} - \frac{2}{27} K^3 \ .
\end{align}
\end{subequations}

Under a Weyl transformation $g_{\mu\nu} \to \tilde{g}_{\mu\nu} = e^{2\sigma} g_{\mu\nu}$, the T-curvatures and the boundary operators transform as
\begin{equation}\label{eq:WeylBT}
\mathcal{B}_{e^{2\sigma}g,k}= e^{-k\sigma}\mathcal{B}_{g,k},\quad \mathcal{T}_{e^{2\sigma}g,k}= e^{-k\sigma}\left(\mathcal{T}_{g,k}+\mathcal{B}_{g,k}\sigma\right),\quad \Phi^{(k)} \to e^{-k\sigma} \left(\Phi^{(k)} - Q \sigma^{(k)}\right) \ \ \text{for } k=0,1,2,3 \ ,
\end{equation}
where we denote
\begin{equation}
\mathcal{B}_{g,k} \sigma|_{\del \mathcal{M}}= \sigma^{(k)} \ .
\label{B}
\end{equation}


\subsection{Boundary Liouville Action}\label{ss:22}
In the presence of a boundary, we have to add boundary terms to the bulk Liouville action in order to have a well defined variation of the action. We would like to construct a classically Weyl invariant action such that fixing the background metric yields a conformal field theory. Every term in the boundary will contain a factor of $\sqrt{\hat{g}}$ coming from the volume element. Under a Weyl transformation,  we have $\sqrt{\hat{g}} \to e^{-3\sigma}\sqrt{\hat{g}} $, hence we should choose the combinations of $\Phi^{(k)}$  (\ref{eq:B}) and $\mathcal{T}_{g,k}$ (\ref{eq:T}) such that under Weyl transformation it receives an overall factor of $e^{3\sigma}$. The Weyl transformation, apart from multiplicative exponential factor, also produces non-homogeneous terms in $\sigma$ (\ref{eq:WeylBT},\ref{B}). We have to make sure that such non-homogeneous pieces are cancelled or vanish when boundary conditions are imposed. 

\subsubsection{Extension of the Action to the Boundary}
We would like to extend the bulk action \eq{eq:bulkac} to the boundary.
Consider first the extension of the 
$\mathcal{Q}$-curvature, whose integral should be a conformal invariant of a manifold with a boundary.
The Gauss-Bonnet-Chern formula for the Euler characteristic of a  four-manifold with boundary reads \cite{CHANG1997327,gonzalez2021eigenvalue}:
\begin{equation}
 \chi(\mathcal{M}) = \f{1}{8\pi^2}\int_{\mathcal{M}}d^{4}x\sqrt{g}\left(\mathcal{Q}_{g}+\frac{1}{4}W^{\mu\nu\sigma\rho}W_{\mu\nu\sigma\rho}\right) + \f{1}{8\pi^2}\int_{\del \mathcal{M}}d^{3}\mathrm{x}\sqrt{\hat{g}}\left(\mathcal{T}_{g,3}-\frac{2}{3}\mathring{K}_{i}^{\; j}\mathring{K}_{j}^{\;k}\mathring{K}_{k}^{\;i}\right) \ .
\end{equation} 
Since the second term in both the bulk and boundary integrals are conformal invariants,  the integral  \begin{equation}\label{eq:clfeuler}
    \int_{\mathcal{M}}d^{4}x\sqrt{g}\ \mathcal{Q}_{g}+ \int_{\del \mathcal{M}}d^{3}y\sqrt{\hat{g}}\ \mathcal{T}_{g,3} \ ,
\end{equation}
is a conformal invariant of ${\cal M}$. Thus, $\mathcal{T}_{g,3}$ is a natural extension of the $\mathcal{Q}$-curvature to the boundary. 

Consider next the kinetic term in the bulk action. In \cite{case2015boundary}, it was shown that
\begin{equation}
   \braket{\phi_1, \phi_2} = \int_{\mathcal{M}}d^{4}x\sqrt{g}\ \phi_1 \mathcal{P}_g \phi_2  +\int_{\del \mathcal{M}}d^{3}\mathrm{x}\sqrt{\hat{g}}\left(\Phi_1^{(0)}\Phi_2^{(3)}+\Phi_1^{(1)}\Phi_2^{(2)} \right) 
\end{equation}
is a symmetric bilinear form and $\braket{\phi, \phi}$ is a conformal invariant. Using above two results, we can extend the bulk Liouville action to the boundary as:
\begin{align}\label{eq:bdryac}
 S_{L}[\phi,g] &= \f{1}{8\pi^2}\int_{\mathcal{M}}d^{4}x\sqrt{g}\left(\phi \mathcal{P}_g \phi+2Q \mathcal{Q}_g\phi+8\pi^2\mu e^{4b\phi}\right)  \nonumber\\
&+ \f{1}{8\pi^2} \int_{\del \mathcal{M}}d^{3}\mathrm{x}\sqrt{\hat{g}}\left(\Phi^{(0)}\Phi^{(3)}+\Phi^{(1)}\Phi^{(2)}+2Q\mathcal{T}_{g,3}\Phi^{(0)}+8\pi^2\mu_Be^{3 b_B\Phi^{(0)}}\right),
\end{align}
where $b_B$ and $\mu_B$ are dimensionless parameters, and the extension of the Liouville potential to the boundary is trivial. The Coulomb gas action ($\mu=0,\  \mu_B=0$) has the Weyl transformation:
\begin{equation}\label{eq:CGWeyl}
S_{CG}[\phi-Q\sigma,e^{2\sigma}g]=S_{CG}[\phi,g] - S_{CG}[Q\sigma,g] - \f{Q}{4\pi^2} \int_{\del \mathcal{M}} dx^3 \sqrt{\hat{g}}\ \left(\Phi^{(1)} - Q\ \sigma^{(1)}\right)\sigma^{(2)} \ .
\end{equation}
We will show later that the last boundary term vanishes upon imposing consistent boundary conditions. Note, that 
the Liouville action is Weyl invariant up to the classical contribution to
the A-type anomaly term. 
If we fix $Q=\frac{1}{b}$ and $b_B = b$, we find that the complete Liouville action \eq{eq:bdryac} transforms as \eq{eq:CGWeyl}. We therefore expect that in the semi-classical limit $b\to 0$, the value of $b_B$ that also ensures the Weyl invariance at the quantum level will satisfy $\lim_{b\to 0} b_B(b) \to b $. 


\subsection{Spectrum and $b_B$}\label{subs:bdyspec}
In the $d$-dimensional Coulomb gas theory, the two point correlation function of $\phi$ in the flat space reads:
\begin{equation}
    \braket{\phi(x)\phi(y)} = \f{2}{d}\log\left(\f{L}{|x-y|}\right) + \cdots \ ,
\end{equation}
where $L$ is an IR regulator. The vertex operators $V_\alpha(x) = e^{d\alpha \phi(x)}$ are conformal primaries with conformal dimensions $\Delta_\alpha = d \alpha (Q-\alpha)$. Conformal invariance dictates that the Liouville interaction term in \eq{eq:bulkac} has
the dimension $\Delta_b =d$, which fixes the quantum corrected background charge to be $Q = b + \f{1}{b}$.
Indeed, in the classical limit  $b \to 0$, we have $Q \approx 1/b$.   

Define boundary vertex operators as $B_{\beta}(\mathbf{x}) = e^{\beta \phi(\mathbf{x})}$.
Consider the theory to be defined on a manifold with topology of a ball\footnote{We can go from a manifold with flat metric and a plate boundary to a 4-ball with an $\mathbb{S}^3$ boundary via a singular Weyl transformation. See appendix \ref{app:singweyl}  } for which $\chi(\mathcal{M}) = 1$. Thus, under a constant Weyl transformation $\sigma(x) = \phi_0$, we have:
\begin{equation}
S_{CG}[\phi+\phi_0, e^{2\phi_0}g] = S_{CG}[\phi,g] + d Q\phi_0 .    
\end{equation}
Invariance of the two point  function of $B_\beta$ under above shift imposes  $Q = 2\beta/d$. Considering the fact that the dimension of $B_\beta$ is zero when $\beta=0$, and the dimension formula should reduce to the known dimension for analogous operator in two dimensions \cite{Fateev:2000ik} fixes the dimension to
\begin{equation}
\Delta_{\beta} = \beta\left(Q-\frac{2}{d}\beta\right) \ .
\end{equation}
Requiring that the boundary Liouville interaction term has the correct dimension i.e. $\Delta_{(d-1)b_B} = d-1$ yields:
\begin{equation}\label{eq:bB}
b_B = \frac{1+b^2-\sqrt{1-b^2(6-8/d)+b^4}}{4(1-1/d)b} \ ,
\end{equation}
where we impose the condition that in the semi-classical limit $b_B = b+O(b^3)$. In $d=2$, formula (\ref{eq:bB})
reduces to expected result $b_B =b$ .

The formula \eq{eq:bB} contains a square root whose argument is negative in a following window 
\begin{equation}
 b \in  \left({\left(-\frac{2 \sqrt{2} \sqrt{d^2-3 d+2}}{d}-\frac{4}{d}+3\right)^{1/2}}, \left({\frac{2 \sqrt{2} \sqrt{d^2-3 d+2}}{d}-\frac{4}{d}+3}\right)^{1/2} \right) \ .
\end{equation}
In this window, $b_B$ is complex. In $d=2$, this window vanishes. In $d=4$, the window is $b \in \left(\left(2-\sqrt{3}\right)^{1/2}, \left(2+\sqrt{3}\right)^{1/2}\right)$ and it asymptotes to $b \in \left(\left(3-2\sqrt{2}\right)^{1/2}, \left(3+2\sqrt{2}\right)^{1/2}\right)$ at large $d$.       
Despite this fact, the dimension of the operator is real and is equal to $d-1$, and hence its correlators are well defined.

\subsection{Conformal Boundary Conditions}
\label{sec:BC}
LCFT in $d$ dimensions is a $d$-derivative theory, where generally, one has to define additional canonical fields and their conjugate momenta \cite{deUrries:1998obu}. The number of additional fields depends on the order of the derivatives in the action, and we can have the usual Neumann and Dirichlet boundary conditions for all of them. A four-derivative theory requires two canonical fields and their conjugate momenta, hence we have four boundary conditions, two for each conjugate pair.

We will study LCFT on the four-dimensional ball, which is conformally equivalent to the upper half space $\mathbb{H}\mathbb{R}^4$. Performing a Weyl transformation (see appendix \ref{app:singweyl}) to the upper half space results in boundary conditions for $\phi$:
\begin{equation}
 \phi(x)= -2Q\log\left(|x|\right)+O(1), \quad  |x|\to\infty  \ .
 \end{equation} 
 Note, that this is the boundary condition away from $y=0$ surface. 
 
\subsubsection{Variation of the Action and Boundary Conditions}
In this subsection, we describe the consistent set of boundary conditions. For an analysis of free theory without background charge see \cite{Chalabi:2022qit}.
We denote $\delta \Phi^{(k)} =0$ for $k = 0,1$ as Dirichlet boundary conditions and  $\delta \Phi^{(3-k)} =0$ for $k = 0,1$ as Neumann boundary conditions. We classify all of them as:
\begin{align}
(D,D) &: \delta \Phi^{(0)} = \delta \Phi^{(1)} =0\qquad
&(D,N)&: \delta \Phi^{(0)} =\delta \Phi^{(2)} =0\nonumber \\
(N,D)&:\delta \Phi^{(3)} =\delta \Phi^{(1)} =0\qquad 
&(N,N)&:\delta \Phi^{(3)} =\delta \Phi^{(2)} =0 \nonumber  \ .
\end{align}
When we impose $\delta \Phi^{(k)} = 0$ for $k>0$ then, unless the variation problem dictates the value of $\Phi^{(k)}$, we are forced to impose $\Phi^{(k)} = 0$ since $\Phi^{(k)}$ carries a non-zero dimension. Also, in order to preserve this boundary condition we 
have to restrict the set of Weyl transformations \eq{eq:WeylBT} and impose $\mathcal{B}_{g,k} \sigma|_{\del \mathcal{M}}= \sigma^{(k)} =0$.  We don't have to impose such a condition when $k=0$ since $\Phi^{(0)}$ is dimensionless and $\Phi^{(0)} = C$, where $C$ is an arbitrary constant, is an allowed boundary condition.

The boundary terms in the variation of the action \eq{eq:bdryac} reads:
\begin{align}
\delta S  &= \frac{1}{4\pi^2}\int_{\del \mathcal{M}} d^{3}\mathrm{x}\sqrt{\hat{g}}\left[\left(\Phi^{(3)}+Q\mathcal{T}_{g,3}+24\pi^2b_B\mu_Be^{3b_B\Phi^{(0)}}\right)\delta\Phi^{(0)} + \Phi^{(2)}\  \delta\Phi^{(1)}\right] \ .
\label{bt}
\end{align}
Using (\ref{bt}) we get following boundary equations for respective boundary conditions:
\begin{subequations}\label{bc}
\begin{align}
(D,D) &:  \Phi^{(1)} = 0,\quad \Phi^{(0)} = C \quad &\implies& \quad \sigma^{(1)}  = 0,  \label{bc:DD}\\
(D,N)&:  \Phi^{(2)} = 0,\quad \Phi^{(0)} = C \quad &\implies& \quad \sigma^{(2)} = 0, \label{bc:DN}\\
(N,D)&:  \Phi^{(1)} = 0,\quad \Phi^{(3)}+Q\mathcal{T}_{g,3}+24\pi^2b_B\mu_B e^{3b_B\Phi^{(0)}}=0  &\implies& \quad  \sigma^{(1)}=0,\label{bc:ND}\\
(N,N)&:  \Phi^{(2)} = 0,\quad \Phi^{(3)}+Q\mathcal{T}_{g,3}+24\pi^2b_B\mu_Be^{3b_B\Phi^{(0)}}=0  &\implies& \quad  \sigma^{(2)}=0 \ . 
\end{align}
\end{subequations}
After imposing boundary conditions (\ref{bc}), we 
see that (\ref{eq:CGWeyl}) becomes: 
\begin{equation}
    S_{CG}[\phi-Q\sigma,e^{2\sigma}g]=S_{CG}[\phi,g] - S_{CG}[Q\sigma,g] \ .
\end{equation}

The boundary conditions simplify when $\mu_B=0$ and restricting to the case of a flat metric with a plate boundary we get:
\begin{subequations}\label{flat_bc}
\begin{align}
(D,D) &:\quad \partial_n\phi|_{\partial \mathcal{M}} = 0, \quad \phi|_{\partial \mathcal{M}} =0,\label{bc:DD_flat}\\
(D,N)&:\quad \partial_n^2\phi|_{\partial \mathcal{M}} = 0,\quad \phi|_{\partial \mathcal{M}} =0,\label{bc:DN_flat}\\
(N,D)&:\quad \partial_n\phi|_{\partial \mathcal{M}} = 0, \quad \partial_n^3\phi|_{\partial \mathcal{M}} =0,\label{bc:ND}\\
(N,N)&:\quad \left( \partial_n^3 + 3 \partial_n \partial^2_i\right )\phi|_{\partial \mathcal{M}} =0, \quad \left(\partial_n^2 - \partial_i^2\right)\phi|_{\partial \mathcal{M}} = 0 \ .\label{bc:NN_flat}
\end{align}
\end{subequations}


\section{Boundary Conformal Anomalies}\label{sec:3}

\subsection{The Stress Tensor and the Displacement Operator}
The trace anomalies of a four-dimensional boundary conformal field theory take the form  \cite{Herzog:2015ioa}: 
\begin{equation}\label{eq:gen-anmly}
\left<T^{\mu}_{\ \mu}\right> = \frac{1}{16\pi^2}\left(c\ W_{\mu \nu\eta\rho}^2-a\ E_4\right)+\frac{\delta(y)}{16\pi^2}\left(a\ E_4^{(\mathrm{bry})}-b_1\ \mathrm{tr}\mathring{K}^3-b_2\ \hat{g}^{\mu\nu}\mathring{K}^{\eta \rho}W_{\mu \nu\eta\rho}\right) \ .
\end{equation}
The $a$ and $c$ anomaly coefficients of the four-dimensional LCFT were computed in \cite{Osborn:2016bev,Levy:2018bdc}.
Our aim is to calculate the $b_1$ and $b_2$ anomaly coefficients.  The $b_1$ and $b_2$ anomaly coefficients can be read from the two- and three-point functions of the displacement operator $D^\nu$ \cite{Herzog_2017,Billo:2016cpy}, defined by
\begin{equation}
    \nabla_\mu T^{\mu\nu}(y,\mathbf{x}) = \delta(y) D^\nu(\mathbf{x})\ .
\end{equation}
In the rest of this section, all the results presented are for flat space with plate boundary, where
only the normal component $D\equiv D^n$ of the displacement operator is non-zero (see Appendix \ref{app:ward_dis}). The displacement operators in this case reads:
\begin{equation}
T^{nn}(y,\mathbf{x})|_{y=0} = D(\mathbf{x}) \ , 
\end{equation}
and its two- and three-point functions take the form \cite{Herzog:2017kkj}:
\begin{equation}\label{eq:anmlydef}
\left<D(\mathbf{x})D(\mathbf{0})\right> = \frac{15\ b_2}{2 \pi^4|\mathbf{x}|^8},\quad \left<D(\mathbf{x})D(\mathbf{x'})D(\mathbf{0})\right> = \frac{35\ b_1}{2 \pi^6|\mathbf{x}|^4|\mathbf{x'}|^4|\mathbf{x-x'}|^4} \ .
\end{equation}

In the higher derivative, $\Delta^2 \phi =0$, free scalar theory without a background charge, 
the stress energy tensor reads
\cite{Osborn:2016bev,Stergiou:2022qqj}: 
\begin{equation}
    8\pi^2 T_{\mu\nu}|_{Q=0} = \frac{4}{3}\partial_{\mu}\partial_{\nu}\partial^{\alpha}\phi\partial_{\alpha}\phi + 4\partial_{\mu}\partial_{\nu}\phi \partial^2\phi - \frac{8}{3}\partial_{\mu}\partial^{\alpha}\phi\partial_{\nu}\partial_{\alpha}\phi - 2\partial_{\mu}\partial^2\phi\partial_{\nu}\phi - 2\partial_{\mu}\phi\partial_{\nu}\partial^2\phi \hspace{4cm} \nonumber \\ \hspace{7.5cm} + \  \delta_{\mu\nu} \left( \f{2}{3} \partial_\alpha\phi \partial^\alpha \partial^2\phi + \f{2}{3} \partial_{\alpha}\partial_{\beta}\phi\partial^{\alpha}\partial^{\beta}\phi - \partial^2\phi\partial^2\phi\right) \ .
    \label{eq:Q0T}
\end{equation}
The contribution to the stress tensor from the background charge $Q$ was computed in \cite{Levy:2018bdc}, and the full stress energy tensor reads:
\begin{equation}
    T_{\mu\nu} = T_{\mu\nu}|_{Q=0} +  \frac{Q}{18\Omega_4} \left( 4\Delta \partial_\mu \partial_\nu -\ \delta_{\mu\nu} \Delta^2 \right) \phi \ .
\end{equation}
  
Let us first only consider the Q-independent expressions.  We begin with rewriting the normal-normal component of bulk stress tensor \eq{eq:Q0T} and separating the derivatives along normal and tangent directions to the boundary ($y=0$):
\begin{equation}
   8\pi^2 D_{0} = \partial^2_y\phi\partial^2_y\phi - 2\partial_y\phi\partial^3_y\phi + 2 \partial_i\partial_y^2\phi \partial_i \phi - \partial_i^2\phi\partial_j^2\phi + 2 \partial_i^2\phi\partial_y^2\phi \hspace{5cm} \nonumber \\ \hspace{5cm} - \f{10}{3} \partial_i^2\partial_y\phi\partial_y\phi + \f{2}{3} \partial_i\partial_j\phi\partial_i\partial_j\phi + \f{2}{3} \partial_i\partial_j^2\phi\partial_i\phi - \f{4}{3} \partial_i\partial_y\phi\partial_i\partial_y\phi \ .  
\end{equation}
Taking the boundary limit of the expression above and plugging (\ref{flat_bc}), we get the form of the displacement operators for all the boundary conditions:
\begin{subequations} \label{D}
    \begin{align}
        8 \pi^2\ D_0^{DD} &= \partial^2_y\phi\partial^2_y\phi,\label{Ddd}\\ 
        8 \pi^2\ D_0^{NN} &= 2 \partial^2_y\phi\partial^2_y\phi - \f{8}{9} \partial^3_y\phi\partial_y\phi + \f{8}{3} \partial_i\partial_y^2\phi\partial_i\phi  + \f{2}{3} \partial_i\partial_j\phi\partial_i\partial_j\phi - \f{4}{3} \partial_i\partial_y\phi\partial_i\partial_y\phi,\label{Dnn}\\
        8 \pi^2\ D_0^{DN} &= -2\partial_y\phi\partial_y^3\phi - \f{10}{3} \partial_i^2\partial_y\phi\partial_y\phi -\f{4}{3} \partial_i\partial_y\phi\partial_i\partial_y\phi,\label{Ddn}\\
        8 \pi^2\ D_0^{ND} &= \partial_y^2\phi\partial_y^2\phi + 2 \partial_i\partial_y^2\phi\partial_i\phi - \partial_i^2\phi\partial_j^2\phi + 2 \partial_i^2\phi\partial_y^2\phi + \f{2}{3} \partial_i\partial_j\phi\partial_i\partial_j\phi + \f{2}{3} \partial_i\partial_j^2\phi \partial_i \phi \ .\label{Dnd}
    \end{align} 
\end{subequations}
With these expressions, we can compute the $b_1$ and $b_2$ anomaly coefficients.

\subsection{Green Function and Boundary Conformal Anomalies}\label{sec:3.2}
The Liouville Green function in four dimensions is a solution to the following equation:
\begin{equation}\label{eq:Gfeq}
    \Delta_x^{2}\ G_0(x, x') = \delta^{(4)}(x-x') \ .
\end{equation}
The solution is not unique as we can add to it the solutions of the homogeneous equation, $\Delta_x^{2}\ F(x, x') = 0$, 
which are important when we  impose  boundary conditions. Denote
by $r^2 = |\textbf{x}-\textbf{x}'|^2 +(y-y')^2$ and $\bar{r}^2 = |\textbf{x}-\textbf{x}'|^2 +(y+y')^2$. The solution to \eq{eq:Gfeq} is
\begin{equation}
   G_0(x, x') = -\f{1}{2}\log\left(\f{r}{L}\right) \ ,
\end{equation}
where $L$ is an IR cutoff and a solution to the homogeneous equation is 
\begin{equation}\label{eq:lgrbsol}
   F(x, x') = -\frac{1}{2}\log\left(\f{\bar{r}}{L}\right) \ .
\end{equation}
Adding this homogeneous solution
\begin{equation}\label{eq:gfnddn}
\braket{\phi(x)\phi(x')} = -\f{1}{2} \log\left(\f{r}{L}\right) + \f{A}{2} \log\left(\f{\bar{r}}{L}\right)
\end{equation}
with $A = +1 $ and $A = -1$, we can satisfy DN and ND boundary conditions respectively. The subgroup of conformal symmetry that preserves the boundary dictates that the Green function in four dimensions should have the form \cite{Chalabi:2022qit}: 
\begin{equation}
    \braket{\phi(x)\phi(x')} = f(\zeta), \quad  \text{with} \quad \zeta = \f{r^2}{4 y y'} \ ,
\end{equation}
where $\zeta$ is the conformally invariant cross ratio. The homogeneous equation then becomes
\begin{equation}
    \zeta^2 (1 + \zeta)^2 f''''(\zeta) + 6\ \zeta (1 + \zeta) (1 + 2\zeta) f'''(\zeta) + 6\ \left(1 + 6 \zeta (1 + \zeta)\right) f''(\zeta) +  12\ (1 + 2 \zeta) f'(\zeta) = 0 \ ,
\end{equation}
where prime is a $\zeta$ derivative. {This is a fourth order ODE, one of its solution is $f=constant$}, and we expect two additional homogeneous solutions apart from the one given above. These solutions are $1/\zeta$ and $1/(1 + \zeta)$. If we consider these power-law solutions along with the log solution \eq{eq:lgrbsol} it is possible to satisfy all the boundary conditions. {We expect the UV behaviour of the greens function to be logarithmic, the $1/\zeta$ solution diverges as power law and hence we ignore it. The $1/(1+\zeta) = 4 yy'/\bar{r}^2$ solution is constant in the UV and is subleading to the logarithmic behaviour. It also vanishes on the boundary ($y=0$) and hence does not contribute to the correlation functions of boundary vertex operator.\footnote{{We thank the referee for pointing this out to us.}} Putting this together, we have 
\begin{equation}\label{eq:gall}
\braket{\phi(x)\phi(x')} = -\f{1}{2} \log\left(\f{r}{L}\right) + \f{A}{2} \log\left(\f{\bar{r}}{L}\right) + B\ \f{y y'}{\bar{r}^2}.
\end{equation}
\noindent Above solution with $(A ,B) = \{(1,1),(-1,-1),(-1,0),(1,0)\}$, satisfies $\{(N, N), (D, D), (D, N), (N, D)\}$ boundary conditions respectively.} 
\subsubsection{Boundary Anomaly Coefficients: Q-independent part}
Using equation \eq{eq:gall} and equations (\ref{D}), we can compute the two- and three-point functions of the displacement operator using Wick's theorem. We get:
\begin{equation}\label{eq:b2anmly}
    \braket{D_0(\textbf{x}) D_0(0)} =\f{2 (2B - A) (A + B) - 2 }{\pi^4 |\textbf{x}|^8} \quad \implies \quad b_2 = \f{4 (2B - A) (A + B) - 4 }{15}  \ ,
\end{equation}
and 
\begin{align}
\braket{D_0(\textbf{x}) D_0(\textbf{x}') D_0(0)} &= -\f{13 + 39 A^2 - 3 (5 + A) (1 + 5 A) B + 30 ( A-1) B^2 + 16 B^3}{9\pi^6 |\textbf{x}|^4|\textbf{x}'|^4|\textbf{x}-\textbf{x}'|^4} \\ \nonumber \implies \quad b_1 &= -2 \times \frac{13 + 39 A^2 - 3 (5 + A) (1 + 5 A) B + 30 (A-1) B^2 + 16 B^3}{315} \ .
\end{align}
The result in \eq{eq:b2anmly} agrees with $d \to 4$ limit of the formulas presented in \cite{Chalabi:2022qit} for general dimensions. Hence at $Q=0$, we have  
\begin{center}
\begin{tabular}{ |c|c|c|c|c| } 
 \hline
 \hspace{1cm} &  \hspace{1cm} &  \hspace{1cm}&  \hspace{1cm}&  \hspace{1cm} \\
 Boundary Condition & NN $(A=1,B=1)$ & DD $(A=-1, B=-1)$ & DN $(A =-1, B=0 )$ & ND $(A=1, B=0)$ \\[3ex]
 \hline
   \hspace{1cm} &  \hspace{1cm} &  \hspace{1cm}&  \hspace{1cm}&  \hspace{1cm} \\
 $b_1$ & $16/63$ & $16/35$ & $-104/315$ & $-104/315$  \\[3ex]
 $b_2$  & $4/15$ & $4/15$& $-8/15$ & $ -8/15$   \\ [3ex]
 \hline
\end{tabular}
\end{center}
\subsubsection{Boundary Anomaly Coefficients: Q-dependent part}
{The Q-dependent part of the displacement operator is 
\begin{equation}
     8 \pi^2 q = \frac{Q}{6} \left( 4\Delta \partial_\mu \partial_\nu -\ \delta_{\mu\nu} \Delta^2 \right) \phi\Bigg|_{y=0} \ = Q \left(\f{1}{2} \partial^4_y\phi + \f{1}{3} \partial^2_i \partial^2_y \phi - \f{1}{6} \partial^2_i \partial^2_j \phi\right)\Bigg|_{y=0}
\end{equation}
such that 
\begin{equation}
    D(\textbf{x}) = D_0(\textbf{x}) +  q(\textbf{x}) \ .
\end{equation}
The Q-dependent contribution to the 2-point function and $b_2$ will come from self contraction of $q(\textbf{x})$ hence we have 
\begin{equation}
    \braket{q(\textbf{x}) q(0)} = -\f{20 B Q^2}{\pi^4 |\textbf{x}|^8}\qquad \implies \qquad \delta b_2 = \f{-40B Q^2}{15}  \ .
\end{equation}
This only contributes to NN and DD boundary conditions. Adding to the result from previous subsection, we have 
\begin{equation}
    \braket{D(\textbf{x}) D(0)} =\f{2 (2B - A) (A + B) - 2 - 20 B Q^2}{\pi^4 |\textbf{x}|^8} \quad \implies \quad b_2 = \f{4}{15} \left((2B - A) (A + B) - 1 - 10 B Q^2\right)  \ .
\end{equation}
In the case of 3-point function, the Q dependent contribution will be 
\begin{equation}
    \braket{D_0(\textbf{x})q(\mathbf{x'})q(0)} + \braket{q(\textbf{x})D_0(\mathbf{x'})q(0)} + \braket{q(\textbf{x})q(\mathbf{x'})D_0(0)} \ .
\end{equation}
We find in our analysis that this quantity is non-zero and non-conformal except for DN boundary condition where $q^{(DN)} = 0$ and gives no contribution. We will leave the resolution of this puzzle for future work.} 

\section{Outlook}\label{sec:5}

Adding boundaries to CFTs often reveals a rather rich structure, and as in the case of two-dimensional
LCFT, it is imperative to study boundary LCFT in higher dimensions. In this work we made first steps
in the study of the four-dimensional case.
Adding boundaries requires a classification of the consistent boundary conditions for these higher derivative theories and
appropriate extensions of the GJMS operators and the ${\cal Q}$-curvature to manifolds with boundary, as done 
in this work.
Judging by the two-dimensional boundary LCFT, the extensions to the boundary of higher-dimensional LCFTs
is likely to reveal new structures such as that of the correlations functions of boundary operators, the conformal boundary anomalies and new duality relations. Another
route to take is the study of line and surface operators that will reveal new aspects of the higher-dimensional LCFTs.
Being nonlocal field theories, the odd-dimensional LCFTs on manifolds with boundary can add another layer of interesting QFT structure.
The stress energy tensor of these theories is nonlocal, and thus the whole realization of the conformal algebra.
This also raises a question about the proper definition and the classification of the anomalies.

RG monotonicity theorems state that certain numbers that characterize the UV and IR fixed points of relativistic quantum field theories
decrease along the flow connecting them. Such numbers are the central charge $c$ of a two-dimensional CFT \cite{Zamolodchikov:1986gt}, the A-type Weyl anomaly
coefficient $a$ of four-dimensional CFTs \cite{Komargodski:2011vj}, the regularized free energy of a three-dimensional CFT on the sphere $F$ \cite{Jafferis_2011} and
the one-dimensional thermal free energy $g$ \cite{PhysRevB.48.7297}.
The proof of typical RG monotonicity theorems assumes unitarity, while 
LCFTs in dimensions higher than two are non-unitary.
Yet, it may still be that such RG constraints can be generalized
also to non-unitary cases, 
and the fact that higher-dimensional LCFTs are tractable makes them a valuable framework to study this question.
In \cite{Chalabi:2022qit}, a rich structure of boundary RG flows is presented by deforming the boundary action with specific relevant boundary operator for specific boundary conditions. It will be valuable to consider the boundary RG flows of higher dimensional
LCFTs.
 
{In this article, we are working in the classical ($b\to 0$) limit. It is possible to include the leading order corrections in $b$. We have obtained anomaly coefficients by
taking boundary limit of the bulk correlators. The Liouville perturbation theory is in the powers of $\mu b^2$ and the Liouville potential is a vertex operator. In perturbation theory, the correction
will contain an additional insertion of the vertex operator in the correlator. We have also briefly mentioned that we find the Q-dependent contribution to the 3 point function of the displacement operator being non-zero and non-conformal except for DN boundary condition where $q^{(DN)} = 0$, and hence gives no contribution. We leave the computation and investigation of these two questions for future work.}

Finally, the original motivation to introducing higher dimensional LCFTs has been for the construction of 
an effective theory for incompressible fluid turbulence  \cite{Oz:2017ihc,oz2020turbulence}.
This proposed application of LCFTs requires the presence of boundaries: one can think of the forcing scale as the boundary of the inertial range of scales where a universal description of turbulence is expected.
Thus, we expect that progress in understanding higher-dimensional boundary LCFTs as initiated in the current work, in particular in the nonlocal three-dimensional case, is likely to provide new insights to the out of equilibrium dynamics of statistical turbulence.

\section{Acknowledgements} 
AG would like to thank S. Biswas for useful discussions.   
This work is supported in part by Israeli Science Foundation excellence center, the US-Israel Binational Science Foundation
and the Israel Ministry of Science.


\begin{appendices}

\section{Ward Identities for Displacement Operator}\label{app:ward_dis}

In this appendix, we will present the ward identities of the displacement operator relevant to us (see \cite{Billo:2016cpy,Miao_2019} for more details). We consider a boundary defined by $y=0$ but not necessarily a flat metric and adopt the notation, $x^\mu = (y,x^i)$. In the Gaussian normal coordinates, we can write metric near the boundary as: 
\begin{equation}
    ds^2 = dy^2 + \left(h_{ij} - 2 y\, K_{ij} + \sum_{n=2}^\infty y^n\, q^{(n)}_{ij}\right) dx^i dx^j \ ,
\end{equation}
where $q^{(n)}_{\mu\nu} = \partial_y^ng^{\mu\nu}/\Gamma(n+1)$. Let $e^\mu_i = \partial X^\mu /\partial x^i$, $h_{\mu\nu} = g_{\mu\nu} + n_\mu n_\nu$ be the transverse metric where $n_\mu$ is the normal, and $\gamma_{ij} = h_{\mu\nu} e^{\mu}_i e^\nu_j$ be the induced metric.\\

Define the generating functional $W = - \text{log}\ Z\left[g_{\mu\nu},X^\mu(x^i)\right]$ where $X^\mu(x^i)$ is general embedding of the boundary, the variation of which reads:
\begin{equation}\label{eq:genvar}
    \delta W =  \f{1}{2} \int dx^d\, \sqrt{g} \braket{T^{\mu\nu}_{bulk}}\, \delta g_{\mu\nu} + \f{1}{2}\int dx^{d-1} \sqrt{h} \, \left[\braket{T_{bdy}^{\mu\nu}} \delta h_{\mu\nu} + 2 \braket{D_\mu} \delta X^\mu + \braket{J^{(n)\mu\nu}}\, \delta q^{(n)}_{\mu\nu}\right] \ . 
\end{equation}
 The full stress tensor is 
$T^{\mu\nu} = T^{\mu\nu}_{bulk} + \delta(y)\, T_{bdy}^{\mu\nu}$.\\ 

\noindent A small reparameterization of the boundary is defined as follows:
\begin{equation}
    \delta g_{\mu\nu} = \delta x^\mu =0, \qquad \delta x^i = - \zeta^i, \qquad \delta X^\mu = \zeta^i\partial_i X^\mu \ .
\end{equation}
This gives 
\begin{equation}
    \delta W = - \int d\textbf{x}^{d-1} \sqrt{h}\ \left[\, \braket{D_\mu} e^{\mu}_i \zeta^i + \cdots \right]\quad \implies \quad \braket{D_\mu} e^{\mu}_i = 0 \ ,
\end{equation}
which means that only non-zero components of $\braket{D_\mu}$ are normal to the boundary.\\

\noindent Invariance under bulk diffeomorphism:
\begin{equation}
    \delta X^\mu  = - \epsilon^\mu, \quad \delta g_{\mu\nu} = \nabla_\mu \epsilon_\nu + \nabla_\nu \epsilon_\mu, \quad \delta n_{\mu} = n^{\alpha}\nabla_{\alpha}\epsilon^\mu, \quad \delta {x}^i =0 \ ,
\end{equation}
implies 
\begin{equation}
    \delta W = - \int dx^d \sqrt{g}\, \epsilon_\mu \nabla_\nu \braket{T^{\mu\nu}_{bulk}} + \int d\textbf{x}^{d-1} \sqrt{h} \left(\braket{T^{\mu\nu}_{bulk}} n_\mu \epsilon_\nu + \braket{T^{\mu\nu}_{bdy}} \f{\delta h_{\mu\nu}}{2} - \epsilon_\mu \braket{D^\mu} + \cdots \right) \ ,
\end{equation}
with $\delta W=0$. Thus, the bulk term immediately gives us 
\begin{equation}
    \nabla_\nu \braket{T^{\mu\nu}_{bulk}} = 0 \ .
\end{equation}
Using $ h_{\mu\nu} = h^{\alpha}_{\mu} h^{\beta}_{\nu}\ g_{\alpha \beta}$ and $h^{\mu}_\nu = \delta^{\mu}_{\nu} + n^{\mu}n_\nu$, we get 
\begin{align*}
    \braket{T^{\mu\nu}_{bdy}} \, \delta h_{\mu\nu} =  \braket{T^{\mu\nu}_{bdy}} \left(h^{\alpha}_{\mu} h^{\beta}_{\nu}\, \delta g_{\mu\nu} + \delta h^{\alpha}_{\mu} h_{\alpha \nu} + \delta h^{\alpha}_{\nu} h_{\alpha \mu}\right)= 2 \braket{T^{\mu\nu}_{bdy}} \left(h_{\nu}^{\alpha} \hat{\nabla}_{\mu} \epsilon_{\alpha} + n_{\mu}h_{\alpha \nu} \nabla_n \epsilon^\alpha \right) \ . 
\end{align*}
Making use of the above derivation, we get (we only look at the boundary terms)
\begin{align}
    \delta W_{bdy} &= \int dy^{d-1} \sqrt{h} \left(\braket{T^{\mu\nu}_{bulk}} n_\mu \epsilon_\nu + \braket{T^{\mu\nu}_{bdy}} \left(h_{\nu}^{\alpha} \hat{\nabla}_{\mu} \epsilon_{\alpha} + n_{\mu}h_{\alpha \nu} \nabla_n \epsilon^\alpha \right) - \epsilon_\mu \braket{D^\mu} + \cdots \right) \ ,
\end{align}
which after integrating by parts becomes
\begin{equation}
    \delta W_{bdy} = \int_{\partial \mathcal{M}} \left(\braket{T^{\mu\nu}_{bulk}} n_\mu \epsilon_\nu - \left(\hat{\nabla}_{\mu} h^\alpha_\nu \right) \braket{T^{\mu\nu}_{bdy}} \epsilon_{\alpha} -   \left(\hat{\nabla}_{\mu}\braket{T^{\mu\nu}_{bdy}}\right) \hat{\epsilon}_{\nu} + \braket{T^{\mu\nu}_{bdy}} n_{\mu} \nabla_n \hat{\epsilon}_\nu  - \epsilon_\mu \braket{D^\mu} + \cdots \right) \ .
\end{equation}
Here we ignored the total derivatives. Using $h_{\mu}^\alpha K_{\alpha \nu} = K_{\mu \nu}$, we can write 
\begin{equation}
 \hat{\nabla}_{\mu} h^\alpha_\nu = K^{\alpha}_\mu n_\nu + K_{\mu \nu} n^\alpha \ . 
\end{equation} 
Using this and imposing $\delta W_{bdy} =0$, we get
\begin{equation}
    \braket{T^{\mu\nu}_{bulk}} n_\mu \epsilon_\nu - \braket{T^{\mu\nu}_{bdy}} K_{\mu \nu} n^\alpha \epsilon_{\alpha} -   \left(\hat{\nabla}_{\mu}\braket{T^{\mu\nu}_{bdy}}\right) h^\alpha_\nu {\epsilon}_\alpha + \braket{T^{\mu\nu}_{bdy}} n_{\nu} \left[ \nabla_n \left(h^\alpha_\mu {\epsilon}_\alpha \right)  + K^\alpha_\mu \epsilon_\alpha \right] - \epsilon_\mu \braket{D^\mu} + \cdots = 0.
\end{equation}
This equation is valid for arbitrary $\epsilon^\mu$, let us choose $\epsilon^\mu = n^\mu$. Then using $n^\alpha K_{\alpha \nu} =0$ and $h^\alpha_\mu {n}_\alpha =0$ yields 
\begin{equation}
    n_\mu \braket{D^\mu} = \braket{T^{\mu\nu}_{bulk}} n_\mu n_{\nu} + \braket{T^{\mu\nu}_{bdy}}\ K_{\mu\nu} + \cdots
\end{equation}
Now we choose $\epsilon^\mu = h^{\mu}_\nu$, which yields  
\begin{align}
    \hat{\nabla}_\mu \braket{T^{\mu\nu}_{bdy}} &= \braket{T^{ \alpha\mu}_{bulk}} n_{\alpha} h_{\mu}^{\nu}\, + \braket{T^{\alpha\mu}_{bdy}} n_{\alpha} K^\nu_\mu  + \cdots
\end{align}
In the case of a flat space with a plate boundary, $K_{\mu\nu} = q^{(n)} = 0$. In this case we can define 
\begin{equation}
    D(y^a) = - T^{\mu\nu}_{bulk}\, n_\mu n_{\nu}  = -T_{yy}(0,y) \ ,
\end{equation}
where $D = D_x = -D^\mu n_\mu$.
\section{Conformal Transformation Between the Ball and  Half Euclidean Space} \label{app:singweyl}
Define a four-dimensional ball
\begin{equation}
   |w|^2 = w_1^2 + w_2^2 + w_3^2 + w_4^2 \leq 1, \qquad \text{with} \qquad ds^2 = dw_1^2 + dw_2^2 + dw_3^2 + dw_4^2  \ ,
\end{equation}
and let the flat four-dimensional half space be defined by coordinates 
\begin{equation}
    -\infty < x^i < \infty,\quad y>0 \qquad i= 1,2,3 \ .
\end{equation}
The coordinate transformation between these two spaces is:
\begin{equation}
   (y,\ {x}^1,\ {x}^2,\ {x}^3,) = \f{1}{w_1^2 + w_2^2 + w_3^2 +(w_4-1)^2 } \ (1 - |w|^2,\ 2w_1,\ 2w_2,\ 2w_3) \ .
\end{equation}
Note, that in the above transformation $|w| \to 1 \implies y \to 0$. The ball metric in the $(y, x^i)$ coordinates reads:
\begin{align}
    g_{\alpha\beta} = e^{2 \Omega} \delta_{\alpha \beta}, \quad \text{with} \quad \Omega &= -\ \text{log}\left( \f{(x^1)^2 + (x^2)^2 + (x^3)^2 + (1+y)^2}{2}\right) = - \text{log}\left(\f{|x|^2 + 2y + 1}{2}\right)\nonumber\\
     &\approx - 2\ \text{log}\ |x| + O(1), \qquad \text{as } |x| \to \infty   \ .
\end{align}

\end{appendices}

\bibliography{draft_1}
\bibliographystyle{JHEP}
\end{document}